**Atomistic potentials for palladium-silver hydrides**

L.M. Hale[1], B.M. Wong[2], J.A. Zimmerman[1], X.W. Zhou[1]

[1]Mechanics of Materials Department, Sandia National Laboratories
P.O. Box 969, Livermore, CA 94550, USA

[2]Materials Chemistry Department, Sandia National Laboratories
 P.O. Box 969, Livermore, CA 94550, USA

Email: lmhale@sandia.gov

PACS:
88.30.rd Inorganic metal hydrides
34.20.Cf  Interatomic potentials and forces
31.15.xv Molecular dynamics and other numerical methods

**Abstract**

New EAM potentials for the ternary palladium-silver-hydrogen system are developed by extending a previously developed palladium-hydrogen potential.  The ternary potentials accurately capture the heat of mixing and structural properties associated with solid solution alloys of palladium-silver.  Stable hydrides are produced with properties that smoothly transition across the compositions.  Additions of silver to palladium are predicted to alter the properties of the hydrides by decreasing the miscibility gap and increasing the likelihood of hydrogen atoms occupying tetrahedral interstitial sites over octahedral interstitial sites.

## 1. Introduction

Considerable work has investigated palladium in regards to its unique hydrogen storage behavior. Palladium is capable of storing a large atomic percent of hydrogen at room temperature and allows for hydrogen to diffuse with a high mobility [1, 2]. While the large atomic mass and cost of palladium limit its use as a storage material for the transportation industry, its unique properties make it ideal for applications where fast diffusion and large storage density are important. One drawback with using palladium is that, as it is for many metals, the addition of hydrogen to a metal adversely affects the material's mechanical properties by reducing ductility and increasing susceptibility to fatigue and corrosion. This is further complicated by the presence of a miscibility gap in the palladium-hydrogen system at room temperature, which results in the separation of the material into a low hydrogen ($\alpha$) phase, and a high hydrogen ($\beta$) phase [2]. There is a large lattice mismatch between the two phases which can result in considerable strain and plastic deformation within the material, and can further increase the likelihood of fatigue and failure [2].

One hydrogen related application where palladium has found a use that requires mechanical resilience is as a hydrogen diffusion membrane [1]. By filtering hydrogen gas through a membrane of palladium, high purity hydrogen can be obtained as the palladium allows for high diffusion of hydrogen atoms while blocking the transmission of impurities. Over time, the membrane is subjected to changing hydrogen pressures, resulting in strain and fatigue. The absorbing and storing of the tritium isotope of hydrogen offers a second example where the mechanical resilience and hydrogen properties are important. Over time, the tritium naturally decays to helium-3 [3-5], which is insoluble in the metal and coalesces into bubbles. The further decay of tritium into helium causes the bubbles to grow creating additional damage and defects within the metal tritide. At a material-specific age, fracturing of the tritide occurs, resulting in rapid release of the helium [3-5].

Alloying palladium with another metal offers the opportunity to alter the properties of the hydride and improve its overall usage. For instance, additions of silver greatly improve the mechanical resilience of palladium hydrides by suppressing the miscibility gap at room temperature [6-8]. Depending on the silver composition, the hydrogen diffusion rate can also be increased with respect to pure palladium [9-11]. However, the addition of silver also decreases the hydrogen solubility [6], resulting in a trade-off of properties with changing compositions.

Atomistic simulations are useful in the evaluation of alloy systems as changes in composition can be more easily explored than with experiments. Here, classical atomistic potentials for the palladium-silver-hydrogen ternary system are developed. The new ternary potentials combine previously developed palladium-hydrogen [12] and silver [13] potentials. Data obtained from quantum mechanical calculations are utilized to fit the cross-interaction functions in the ternary potentials. The properties predicted by the potentials are calculated for a large range of compositions. The simulated materials show smooth changes in the lattice parameter, cohesive energy and bulk modulus with varying composition. Solid solutions are predicted for interactions between palladium and silver, with the structural and energetic properties consistent with experimental results. The simulated hydrides give a qualitative representation of a number of key material behaviors, including a shift in the hydrogen site occupancy and the disappearance of the miscibility gap at 300 K with increasing silver.

## 2. Potential Format

The ternary potential is constructed as an embedded atom method (EAM) potential [14-16]. EAM potentials are selected here to model palladium silver hydrides as this class of potentials is computationally inexpensive and well suited for modeling face-centered-cubic metals and alloys. While more complex hydrides [17, 18] would likely require more computationally intensive potentials, the structure of palladium and palladium-silver hydrides can be modeled with EAM as the crystal structure is a face-centered-cubic metal lattice with hydrogen interstitials. This is supported by the fact that the pre-existing EAM palladium hydride potential by Zhou [12] is capable of modeling not only the structure but many of the materials properties as well.

The general form of an EAM potential is given by

$$E_C = \frac{1}{N}\left[\sum_{i=1}^{N} F_i(\bar{\rho}_i) + \frac{1}{2}\sum_{i=1}^{N}\sum_{\substack{j=1 \\ j \neq i}}^{N} \phi_{ij}(r_{ij})\right] \quad (1)$$

$$\bar{\rho}_i = \sum_{\substack{i=1 \\ j \neq i}}^{N} \rho_j(r_{ij}). \quad (2)$$

Here, $E_c$ is the cohesive energy, $N$ is the total number of atoms, $r_{ij}$ is the atomic spacing between atoms $i$ and $j$, $F_i$ is the embedding energy function for atom $i$, $\rho_j$ is the electron density function for atom $j$, and $\phi_{ij}$ is the pair interaction function between atoms $i$ and $j$. $\bar{\rho}_i$ is the total electron density felt by atom $i$ from all other atoms $j$. EAM places no limitations on the exact mathematical expressions used for the three functions $F$, $\rho$, and $\phi$, but practice and theory point to particular characteristic forms for each.

For single element systems, a specific EAM potential is not unique and certain transformations of the functions can change their mathematical form without altering the properties of the element [19]. The total energy is a linear expression of the embedding energy and pair functions, allowing them to be transformed according to

$$F_i^{Final}(\bar{\rho}_i) = F_i^{Initial}(\bar{\rho}_i) + k\bar{\rho}_i \quad (3)$$

$$\phi_{ij}^{Final}(r_{ij}) = \phi_{ij}^{Initial}(r_{ij}) - 2k\rho(r_{ij}), \quad (4)$$

where $k$ is an arbitrary constant. The initial potential given by embedding function $F_i^{Initial}$ and pair function $\phi_{ij}^{Initial}$ is equivalent to the final potential given by embedding function $F_i^{Final}$ and pair function $\phi_{ij}^{Final}$. Additionally, the units of the electron density are arbitrary and can be freely scaled due to the fact that it only arises within the embedding function.

When constructing an alloy EAM system, an additional pair function is required for each unique atom type pairing. The alloy pair function is highly dependent on and specific to the elemental functions used. As the embedding energy and electron density functions are defined for elements only, the role of the transformation and scaling of these functions greatly impacts the alloy interaction. Thus, in developing an alloy system, specifications must be made for selecting or fitting the transformation constant, $k$, and defining the relative magnitudes of the electron density associated with the different atoms. While $k$ can be treated as a fitting parameter, the previous work by Zhou, et al. [12] showed that it is not fully independent of the other parameters and can drift to unphysical values if left unconstrained. Here, we follow the work of Zhou, et al. [12] and set $k$ to the unique value which results in both the embedding and pair functions being independently minimized for the ideal face centered cubic structure and lattice spacing.

Successfully creating an atomic potential for the Pd-Ag-H ternary system using EAM requires a total of twelve functions: three embedding energy functions, $F_{Pd}$, $F_{Ag}$, and $F_H$, three electron density functions, $\rho_{Pd}$, $\rho_{Ag}$ and $\rho_H$, and six pair functions, $\phi_{Pd-Pd}$, $\phi_{Ag-Ag}$, $\phi_{H-H}$, $\phi_{Pd-Ag}$, $\phi_{Pd-H}$ and $\phi_{Ag-H}$. As opposed to developing new formulations of all twelve functions, the procedure here utilizes existing base potentials to form a framework for building up to the full system. By combining the existing Pd-H potential previously developed by this group [12] with a published Ag potential, only new functions for $\phi_{Pd-Ag}$ and $\phi_{Ag-H}$ are necessary.

### 3. Quantum Calculations

To obtain a fitting basis, values for the lattice parameters, cohesive energies, and bulk moduli are obtained from density functional theory (DFT) using the VASP code. The DFT results are based on generalized gradient approximation (GGA) methods using Projector-Augmented-Wave (PAW) pseudopotentials with a dispersion-corrected Perdew-Burke-Ernzerhof (PBE-D2) functional [20]. Recent work [21] has shown that the use of dispersion corrections is important for obtaining improved cohesive energies and lattice constants of metals and semiconductors. Within the DFT-D2 approach [22, 23], an atomic pair-wise dispersion correction is added to the Kohn-Sham part of the total energy (EKS-DFT) as

$$E_{\text{DFT-D}} = E_{\text{KS-DFT}} + E_{\text{disp}}, \qquad (5)$$

where $E_{\text{disp}}$ is given by

$$E_{\text{disp}} = -s_6 \sum_{i=1}^{N-1} \sum_{j=i+1}^{N} \sum_{\mathbf{g}} f_{\text{damp}}\left(R_{ij,\mathbf{g}}\right) \frac{C_6^{ij}}{R_{ij,\mathbf{g}}^6}. \qquad (6)$$

Here, the summation is over all atom pairs $i$ and $j$, and over all $\mathbf{g}$ lattice vectors with the exclusion of the $i = j$ contribution when $\mathbf{g} = 0$ (this restriction prevents atomic self-interaction in the reference cell). The parameter $C_6^{ij}$ is the dispersion coefficient for atom pairs $i$ and $j$, calculated as the geometric mean of the atomic dispersion coefficients: $C_6^{ij} = \sqrt{C_6^i C_6^j}$. The $s_6$ parameter is a global scaling factor which is specific to the adopted DFT method ($s_6 = 0.75$ for PBE), and $R_{ij,\mathbf{g}}$ is the interatomic distance between atom $i$ in the reference cell and $j$ in the neighboring cell at distance $|\mathbf{g}|$. A cutoff distance of 30.0 Å is used to truncate the lattice summation. In order to avoid near-singularities for small interatomic distances, the damping function has the form

$$f_{\text{damp}}\left(R_{ij,\mathbf{g}}\right) = \frac{1}{1 + \exp\left[-d\left(R_{ij,\mathbf{g}}/R_{\text{vdW}} - 1\right)\right]}, \qquad (7)$$

where $R_{vdW}$ is the sum of atomic van der Waals radii $\left(R_{vdW} = R_{vdW}^{i} + R_{vdW}^{j}\right)$, and $d$ controls the steepness of the damping function.

The systems investigated are constructed as $3 \times 3 \times 3$ metal supercells in which silver atoms are randomly substituted for palladium atoms to produce $Pd_{1-x}Ag_x$ alloy concentrations of approximately x = 0, 0.2, 0.25, 0.3, 0.35, 0.4, and 1. For these calculations, a cutoff energy of 350 eV for the plane-wave basis set is used. The Brillouin zone is sampled using a $2 \times 2 \times 2$ Gamma-centered Monkhorst-Pack grid. The system and mesh sizes are selected such that both the alloys and hydrides could be handled with the same supercell size to allow for direct energy comparisons. K-point convergence was obtained for PdH using this size. Larger k-point meshes are currently too computationally expensive for the fully saturated hydrides. Optimization of both the nuclear coordinates and cell parameters are carried out simultaneously for the ions and the unit cell.

Additional information for fitting the $\phi_{Ag-H}$ pair function is obtained from idealized systems of $Pd_{0.75}Ag_{0.25}H_y$. The calculations consist of single unit cell arrangements containing four metal atoms and various amounts of hydrogen placed at either octahedral or tetrahedral interstitial sites. The periodic nature of the system results in only three unique interstitial sites: O1, an octahedral site surrounded by 6 palladium atoms, O2, an octahedral site surrounded by 4 palladium and 2 silver atoms, and T, the tetrahedral site surrounded by 3 palladium and 1 silver atoms (figure 1). The small system size allows for the use of a very high plane-wave basis set cutoff energy of 500 eV, and a dense $15 \times 15 \times 15$ Gamma-centered Monkhorst-Pack grid.

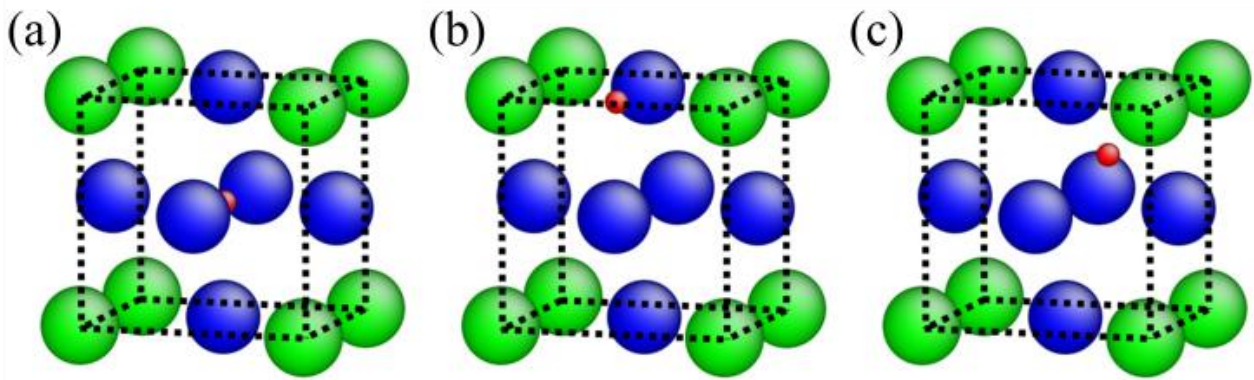

**Figure 1. Diagrams showing the three unique interstitial sites in the idealized Pd0.75Ag0.25 lattice. (a) O1 octahedral site where the hydrogen (small red) atom is surrounded by 6 palladium (face-centered blue) neighbors. (b) O2 octahedral site where the hydrogen is surrounded by 2 silver (corner, green) and 4 palladium neighbors. (c) T tetrahedral site where the hydrogen is neighbored by 1 silver (top front right corner) and 3 palladium atoms (top, front and right face centers).**

4. **Optimization Approach**

The development and fitting of the $\phi_{Pd-Ag}$ and $\phi_{Ag-H}$ pair functions is accomplished using Mathematica [24] in conjunction with the LAMMPS molecular dynamics software [25, 26]. The two pair functions are optimized separately, with the parameters for $\phi_{Pd-Ag}$ being obtained first. $\phi_{Ag-H}$ is developed with the end goal of modeling palladium rich $Pd_{1-x}Ag_xH_y$ compounds, and therefore is dependent on $\phi_{Pd\text{-}Ag}$.

First, a number of initial parameter sets for each pair function are selected. These initial sets are selected such that the $\phi_{Pd\text{-}Ag}$ pair function approximates the $\phi_{Pd\text{-}Pd}$ and $\phi_{Ag\text{-}Ag}$ pair functions, and the $\phi_{Ag\text{-}H}$ pair function is close to the $\phi_{Pd\text{-}H}$ pair function. Mathematica constructs an EAM potential table for each of the parameter sets.

Next, each of the potential tables is used in LAMMPS to calculate the predicted properties for a number of test systems. The test systems are all $3 \times 3 \times 3$ unit cells and are constructed from the relaxed atomic coordinates obtained from the quantum calculations. Each system is adjusted by changing the dimensions to match scaled values of the lattice parameters. The scaled lattice parameters are obtained by taking the values calculated from DFT and linearly scaling them with respect to composition such that the values of palladium hydrides and pure silver match the EAM predictions. This is done as these compositions are not affected by the new pair functions and therefore their associated predicted properties will not change.

The results of the molecular dynamics simulations are then read back into Mathematica, where a goodness of fit, $Q$, is evaluated as

$$Q = \sum_{i=1}^{Np} \sum_{j=1}^{Ns} w_i \left( \xi_{ij} - \xi_{ij}^0 \right)^2 . \tag{8}$$

Here, $Np$ is the number of different properties being fitted, $Ns$ is the number of systems evaluated, $w_i$ is a weighting constant for each property, and $\xi_{ij}$ and $\xi_{ij}^0$ are the measured and targeted values for each property and system. Values for $w_i$ are selected such that the contributions of all the properties to the goodness of fit are of the same order of magnitude. The goodness of fit value is calculated for the different parameter sets and used with a Nelder-Mead minimization algorithm [27] to iterate and find improved parameter sets. Iterations cease when the material properties calculated for the parameter sets converge.

$\phi_{Pd\text{-}Ag}$ is determined by calculating the system pressures, interatomic forces, bulk modulus and cohesive energy values for $Pd_{1-x}Ag_x$ alloys. The pressures and forces are targeted to zero allowing structural information, including the lattice parameters, to influence the fit. Target values for the cohesive energy are selected to match the experimental heat of mixing [28] in order to insure the favorability of mixing. The bulk modulus target values are obtained by scaling the DFT values in a similar fashion as is done for the scaled lattice parameters.

$\phi_{Ag\text{-}H}$ is optimized by utilizing $Pd_{0.75}Ag_{0.25}H_y$ compounds and calculating the system pressures, interatomic forces and cohesive energy values. The cohesive energy target values are scaled from the DFT results such that the target and measured energies of $Pd_{0.75}Ag_{0.25}$ are identical. Bulk modulus values are not included as the systems used are highly asymmetric.

5. **Potential Functions**
   5.1.   $F_{Pd}$, $F_H$, $\rho_{Pd}$, $\rho_H$, $\phi_{Pd\text{-}Pd}$, $\phi_{H\text{-}H}$, and $\phi_{Pd\text{-}H}$

The interactions for and between palladium and hydrogen are modeled using the Pd-H potential by Zhou, et al. [12]. The Zhou potential is notable for representing the entire hydrogen composition range and predict the miscibility gap associated with adding hydrogen to palladium [29, 30]. Additionally, it correctly predicts hydrogen occupying octahedral interstitial positions, gives good elastic constants, and shows a decrease in the tensile strength with increasing hydrogen.

This potential was constructed by adding hydrogen interactions to a palladium potential developed by Foiles and Hoyt [31]. The hydrogen elemental functions and $\phi_{Pd-H}$ use the formulas given in Zhou, et al. [12]. An alternative formulation is used for the palladium elemental functions given by the high-order polynomial functions

$$\phi_{Pd-Pd} = \sum_{n=1}^{12} B_n * (r_{ij} - r_c)^n \qquad (9)$$

$$\rho_{Pd} = \sum_{n=1}^{18} C_n * (r_{ij} - r_c)^n \qquad (10)$$

$$F_{Pd} = \sum_{n=1}^{14} D_n * (\bar{\rho})^n. \qquad (11)$$

Here, $B_n$, $C_n$, and $D_n$ are fitting constants, which are given in table 1. These functions are transformations of the functions presented in Zhou, et al. [12] and only differ numerically due to the precision of the constants. The palladium functions presented here are used due to their simpler form, and the fact that $\rho_{Pd}$ and $\phi_{Pd-Pd}$ are exactly zero at the cutoff distance of $r_c = 5.35$ Å.

**Table 1. The fitting coefficients for the palladium elemental potential, as given by equations (9-11). All of the coefficients are in units corresponding to $F_{Pd}$ and $\phi_{Pd-Pd}$ in eV, and $r_{ij}$ in Å.**

| | |
|---|---|
| $B_1 = -0.028373766572053864$ | $B_7 = -5.771864104430561$ |
| $B_2 = -0.693912140600952$ | $B_8 = -2.28830864482984$ |
| $B_3 = -2.7016415915398144$ | $B_9 = -0.5877004782115718$ |
| $B_4 = -6.7769673718754095$ | $B_{10} = -0.09383206080420499$ |
| $B_5 = -10.162640526793803$ | $B_{11} = -0.008435154429755613$ |
| $B_6 = -9.52751913023066$ | $B_{12} = -0.00032528471668893122$ |
| | |
| $C_1 = 0.002060374409708691$ | $C_{10} = -98.24540107806739$ |
| $C_2 = -0.0023415161219090506$ | $C_{11} = -41.92138213965177$ |
| $C_3 = -2.41367586430302$ | $C_{12} = -13.417138960520848$ |
| $C_4 = -17.06385814025267$ | $C_{13} = -3.2018945664819363$ |
| $C_5 = -62.34712859222751$ | $C_{14} = -0.5606757870370728$ |
| $C_6 = -140.20094946285158$ | $C_{15} = -0.06986620891831186$ |
| $C_7 = -211.12596830146109$ | $C_{16} = -0.005858811498577358$ |
| $C_8 = -223.7860893116957$ | $C_{17} = -0.0002961001445449645$ |
| $C_9 = -172.29613294405436$ | $C_{18} = -6.808692841119764*10^{-6}$ |
| | |
| $D_1 = 0.4033050646864627$ | $D_8 = 4.4667023611152954*10^{-7}$ |
| $D_2 = -0.08376556579117103$ | $D_9 = -1.2956762121440043*10^{-8}$ |
| $D_3 = -0.03472823555191091$ | $D_{10} = 2.6544659789031297*10^{-10}$ |
| $D_4 = 0.013546315882438402$ | $D_{11} = -3.762668569352921*10^{-12}$ |
| $D_5 = -0.0020563045380577022$ | $D_{12} = 3.51436199733632*10^{-14}$ |
| $D_6 = 0.00018399325080912537$ | $D_{13} = -1.946895226210271 * 10^{-16}$ |
| $D_7 = -0.000010884705301682515$ | $D_{14} = 4.847680017321573 * 10^{-19}$ |

### 5.2.   $F_{Ag}$, $\rho_{Ag}$, and $\phi_{Ag\text{-}Ag}$

Rather than fitting new $F_{Ag}$, $\rho_{Ag}$, and $\phi_{Ag\text{-}Ag}$ functions, four previously published silver potentials are examined as possible candidates. For simplicity, the four silver potentials investigated are referred to using the leading author's last name: Adams [32], Zhou [33], Williams [13] and Sheng [34].

Prior to evaluating, the four silver potentials were normalized according to the scheme used in the development of the PdH$_y$ potential [12]. First, $F_{Ag}$, and $\phi_{Ag\text{-}Ag}$ are transformed using equations (3) and (4) to the unique value of $k$ where both functions are minimized for the equilibrium zero pressure face-centered-cubic structure. This transformation is useful as it allows for the functions of the different potentials to be directly compared and gives them a physically intuitive shape. For further comparison, $\rho_{Ag}$ is scaled such that the total electron density felt by an atom $\bar{\rho}_i$ equals 1 for the equilibrium zero pressure face-centered cubic lattice.

The normalized EAM functions for the four silver potentials are shown in figure 2. Even though the numerical models used to develop these functions are distinctly different for the four potentials, they are observed to have comparable forms within the effective range of the potentials after normalization. The similarity of the normalized curves offers confidence in the normalization method used.

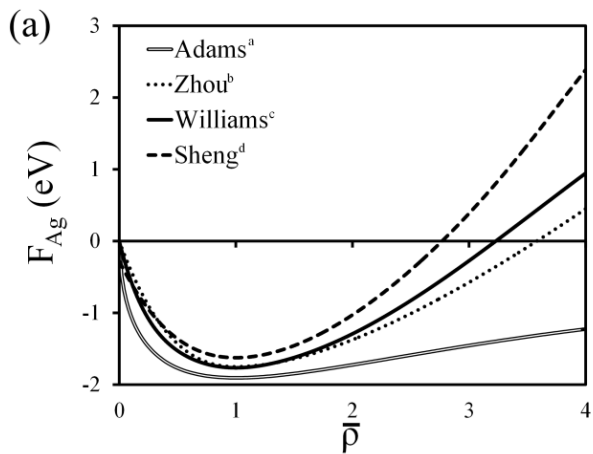

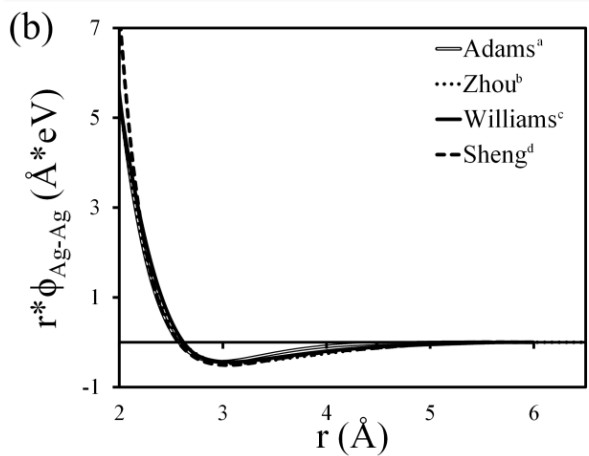

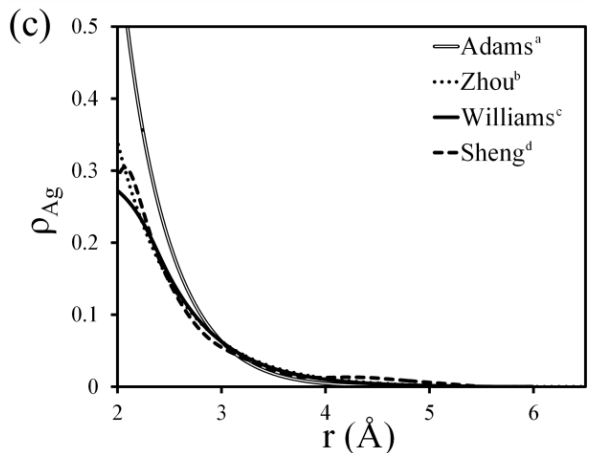

[a]Reference [32]
[b]Reference [33]
[c]Reference [13]
[d]Reference [34]

**Figure 2.** The normalized potential functions for (a) embedding energy, $F_{Ag}$, (b) electron density, $\rho_{Ag}$, and (c) pair potential, $r*\phi_{Ag-Ag}$, for the silver potentials investigated.

Table 2 contains structural and elastic properties, along with the energies for various defects calculated for the different silver potentials. While all of the potentials provide excellent structural, lattice, and elastic properties, the Sheng [34] and Williams [13] potentials excel over the Adams [32] and Zhou [35] silver potentials with respect to the energy of planar defects. In the end, the Williams potential was selected over the Sheng potential as the member functions of the Williams potential have explicit forms, thus making the Williams potential more amenable to the fitting process.

**Table 2. Comparison of various properties calculated with the different EAM silver potentials to reported experimental and ab initio values.**

| Property | Experiment | Ab initio | Adams | Zhou | Williams | Sheng |
|---|---|---|---|---|---|---|
| **Structure** | | | | | | |
| $E_0$ (eV) | -2.85 [13] | | -2.848 | -2.850 | -2.85 | -2.852 |
| $a_0$ (Å) 0 K | | | 4.0907 | 4.0896 | 4.09 | 4.0638 |
| $a_0$ (Å) 300 K | 4.09 [36] | | 4.1147 | 4.1188 | 4.1146 | 4.0883 |
| **Elastic (GPa)** | | | | | | |
| B 0 K | 108.7 [37] | 115.17 [38] | 104 | 103.6 | 104 | 109.9 |
| C11 0 K | 131.5 [37] | 132.99 [38] | 129 | 124 | 124.2 | 131.7 |
| C12 0 K | 97.3 [37] | 106.26 [38] | 91 | 93.7 | 93.9 | 99.2 |
| C44 0 K | 51.1 [37] | 61.38 [38] | 56.7 | 46.2 | 46.4 | 51.2 |
| B 300 K | 103.8 [37] | | 92.7 | 92.2 | 100.4 | 101.7 |
| C11 300 K | 124.0 [37] | | 114.6 | 107.4 | 121.8 | 122.8 |
| C12 300 K | 93.7 [37] | | 81.8 | 82.8 | 90 | 92.5 |
| C44 300 K | 46.1 [37] | | 50.1 | 38.3 | 46.7 | 45.9 |
| **Vacancy** | | | | | | |
| $E_f$ (eV) | 1.1 [39] | 1.06-1.20 [40, 41] | 0.97 | 1.16 | 1.14 | 1.22 |
| **Planar Defects (mJ/m$^2$)** | | | | | | |
| $\gamma_{\text{Intrinsic Stacking Fault}}$ | 16-20 [42] | 21-50 [43-46] | 1.5 | 5.96 | 17.8 | 26.1 |
| $\gamma_{\text{Unstable Stacking Fault}}$ | | 95-190 [45, 46] | 118.5 | 89.9 | 114.8 | 107.9 |
| $\gamma_{\text{Twin}}$ | 8 [42] | 14-19 [43, 44] | 0.75 | 3.57 | 9.15 | 13.11 |
| **Surface Energies (mJ/m$^2$)** | | | | | | |
| (110) | 1140-1246 [42] | 1260-1400 [47, 48] | 764 | 1083 | 1017 | 1118 |
| (100) | 1140-1246 [42] | 1210-1300 [47, 48] | 702 | 962 | 941 | 1038 |
| (111) | 1140-1246 [42] | 1210 [40, 47] | 632 | 925 | 877 | 960 |

*5.3. $\phi_{Pd-Ag}$*

Two alternate models for the $\phi_{Pd-Ag}$ pair functions are simultaneously developed and fitted. The first model is a generalized Morse function of the form

$$\phi(r) = \{\beta * D * \exp[-\alpha(r-r_0)] - \alpha * D * \exp[-\beta(r-r_0)]\}\psi(r), \quad (12)$$

where $\alpha$, $\beta$, $D$, and $r_0$ are constant parameters and $\psi$ is the cutoff function

$$\psi(r) = \left| \begin{array}{ll} 1 & r < (r_c - r_s) \\ \frac{1}{2}\left\{1 + \cos\left[\frac{\pi*(r - r_c + r_s)}{r_s}\right]\right\} & (r_c - r_s) \leq r \leq r_c , \\ 0 & r > r_c \end{array} \right. \qquad (13)$$

which contains the additional parameters of $r_c$ and $r_s$. Five parameters are included in the fitting of the Morse function: $\alpha$, $\beta$, $D$, $r_0$ and $\bar{\rho}_{Ag}/\bar{\rho}_{Pd}$. $\bar{\rho}_{Ag}/\bar{\rho}_{Pd}$ is a constant that scales the normalized $\rho_{Ag}$ function with respect to the normalized $\rho_{Pd}$ function for the cross interactions. Values for $r_c$ and $r_s$ are chosen to provide a smooth cutoff.

The second model is a hybridization of the elemental pair functions for silver and palladium. The idea for this model is to use the information from the element pair functions to create an alloy pair function instead of fitting to an arbitrary model. This approach seems justified here as palladium and silver are completely solid soluble in each other and the shape of the elemental pair functions are close to each other. The alloy pair function is constructed by a linear combination of the elemental pair functions:

$$\phi_{Pd-Ag}(r) = M_1 * \phi_{Pd-Pd}(r) + M_2 * \phi_{Ag-Ag}(r). \qquad (14)$$

Note that the weighting parameters $M_1$ and $M_2$ are treated independently. The hybrid model has the distinct advantage over the Morse model in that a smooth function is created by fitting only three parameters: $M_1$, $M_2$, and $\bar{\rho}_{Ag}/\bar{\rho}_{Pd}$.

For clarity, the potentials developed from the Morse $\phi_{Pd-Ag}$ pair function and the hybridized $\phi_{Pd-Ag}$ pair function are subsequently referred to as potential m and potential h, respectively. The fitted parameters for the two $\phi_{Pd-Ag}$ pair functions are listed in table 3. Plots of the elemental and alloy pair functions are shown in figure 3(a).

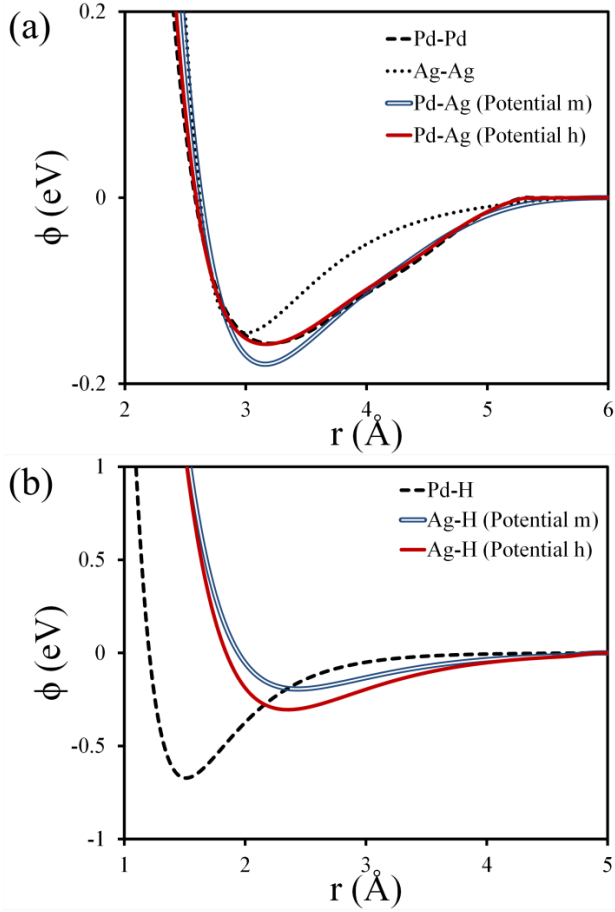

**Figure 3. Plots of the pair functions. (a) The elemental and metal alloy pair functions. (b) The metal-hydrogen pair functions.**

### 5.4. $\phi_{Ag\text{-}H}$

Two alternate $\phi_{Ag\text{-}H}$ pair functions are constructed: one for potential m and one for potential h. Both $\phi_{Ag\text{-}H}$ pair functions are modeled using a generalized Morse function, of the same form given in equation (13). The fitted parameters for the two potentials are given in table 3. The $\phi_{Ag\text{-}H}$ pair function curves are shown alongside the $\phi_{Pd\text{-}H}$ pair function from Zhou [12] in figure 3(b).

**Table 3. The fitted parameters for the two EAM potentials.**

|  | Potential m | Potential h |
|---|---|---|
| $\alpha_{Pd\text{-}Ag}$ | 2.733203 | |
| $\beta_{Pd\text{-}Ag}$ | 1.239651 | |
| $D_{Pd\text{-}Ag}$ | 0.119856 | |
| $r_{0\ Pd\text{-}Ag}$ (Å) | 3.160349 | |
| $r_{c\ Pd\text{-}Ag}$ (Å) | 6 | |
| $r_{s\ Pd\text{-}Ag}$ (Å) | 1.8 | |
| $M_1$ | | 0.888823 |
| $M_2$ | | 0.136682 |
| $\bar{\rho}_{Ag}/\bar{\rho}_{Pd}$ | 0.744514 | 1.002333 |
| $\alpha_{Ag\text{-}H}$ | 2.49019 | 2.16562 |
| $\beta_{Ag\text{-}H}$ | 1.64021 | 1.76185 |
| $D_{Ag\text{-}H}$ | 0.227851 | 0.755217 |
| $r_{0\ Ag\text{-}H}$ (Å) | 2.43479 | 2.35349 |
| $r_{c\ Ag\text{-}H}$ (Å) | 4.95 | 4.95 |
| $r_{s\ Ag\text{-}H}$ (Å) | 0.3 | 0.3 |

## 6. Characteristics of Palladium Silver Alloys

The properties of the $Pd_{1-x}Ag_x$ alloys are evaluated by performing molecular dynamics simulations on a series of systems containing 4000 metal atoms ($10 \times 10 \times 10$ unit cells). The alloy systems are constructed by randomly substituting silver atoms for palladium atoms in concentrations of every 0.02 from x = 0 to 0.50. The systems are annealed at 300 K and zero pressure using NPT integration for 50 picoseconds, before quenching to 0 K over 10 picoseconds and then relaxing the system by minimizing the atomic energy and forces.

Plots of the equilibrium lattice constant and cohesive energy at 0 K as a function of x are shown in figure 4. For comparison, the fitting data used and published experimental values are included in the figures. Both EAM potentials give an excellent agreement with the fitting targets, with the lattice constants matching within 0.1% and the cohesive energies matching within 0.5%. Figure 5 compares the heat of mixing predicted by the EAM potentials to experimental and quantum values. As previously stated, the EAM potentials are fitted to the experimental heat of mixing. While the heat of mixing predicted by EAM does not fully capture the experimentally observed behavior, the magnitude of the EAM values are closer to the experimental values than the DFT values calculated here.

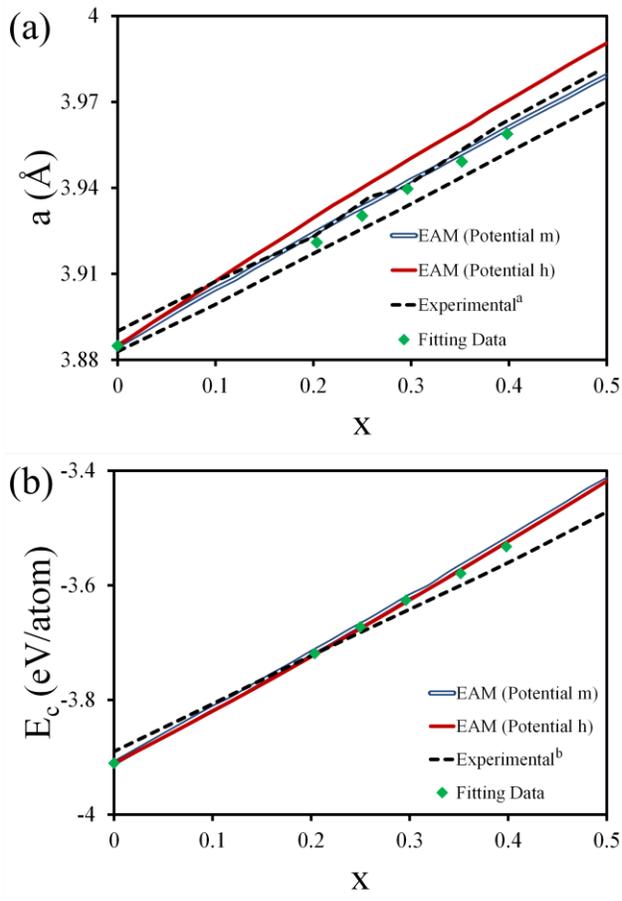

[a]References [7, 49, 50]
[b]Reference [28]

**Figure 4. (a) Lattice parameters, a, and (b) cohesive energies, $E_c$, calculated and reported for $Pd_{1-x}Ag_x$ alloys.**

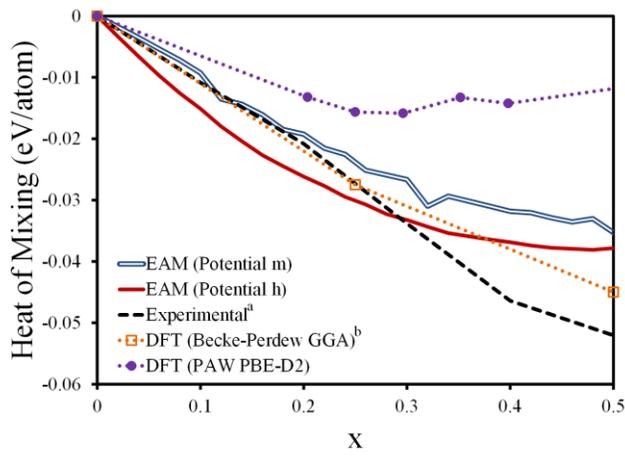

[a]Reference [28]
[b]Reference [51]

**Figure 5. A comparison of the heat of mixing of $Pd_{1-x}Ag_x$ alloys obtained from EAM, DFT and experimental measurements.**

Elastic properties have also been evaluated. Figure 6 shows the values obtained for the 0 K bulk modulus, B, and the single crystal elastic constants C11, C12 and C44. Values for the bulk modulus are within 4.4% of the target fitting values. As the fitting values are scaled to match the properties of the elemental EAM potentials, the change in the bulk modulus with composition is consistent with quantum calculations and the short range of experimental results. Similar matching trends are also observed for C11 and C12. The EAM palladium potential under predicts C44 in comparison with both the experimental and DFT results, which results in C44 also being lower for the alloys. Evaluating the composition based trend of C44 proves murky as DFT shows C44 to always decrease with Ag additions, while the experimental results show an increase in C44, at least for the limited composition range explored.

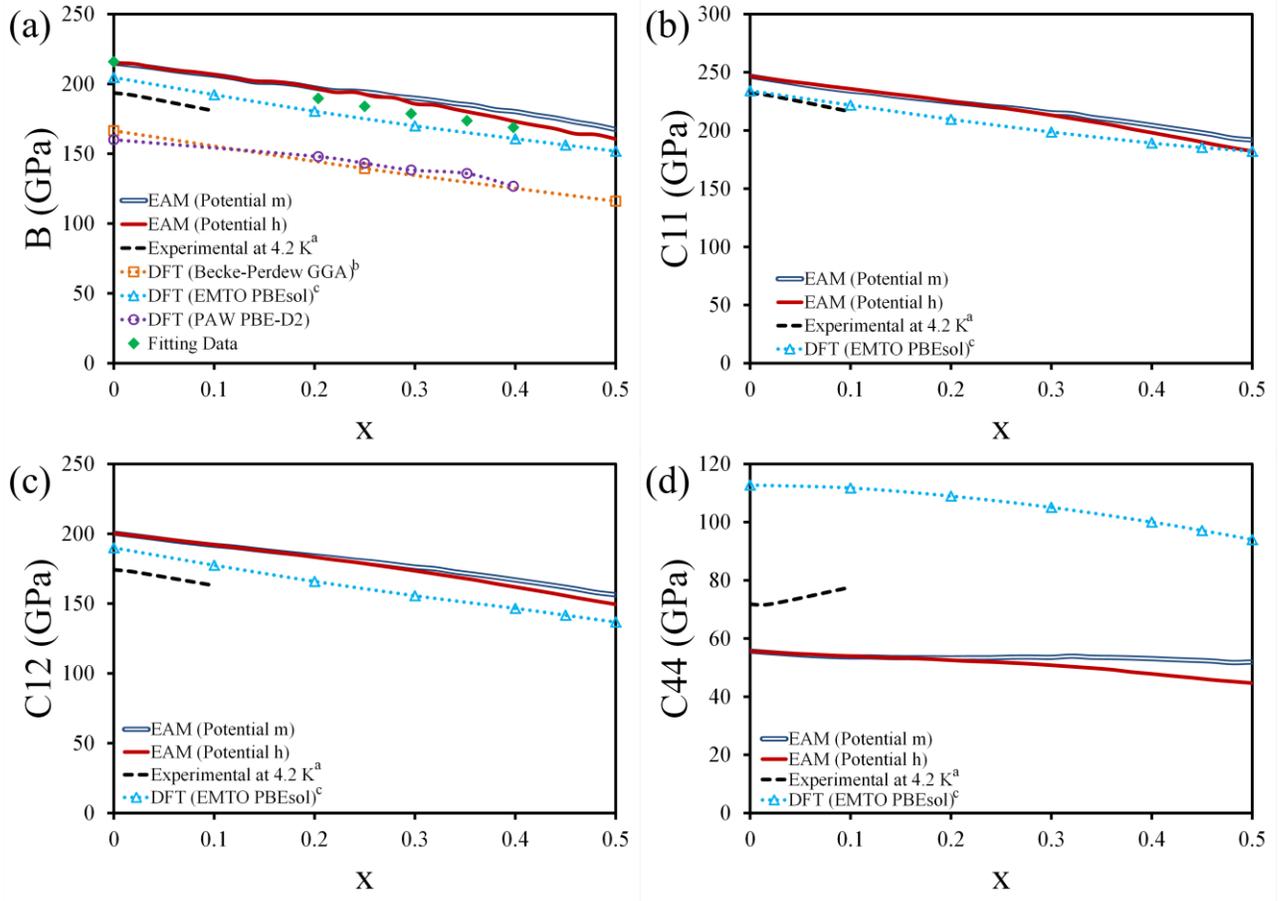

[a]Reference [52]
[b]Reference [51]
[c]Reference [38]

**Figure 6. Plots of the elastic constants (a) bulk modulus, B, (b) C11, (c) C12, and (d) C44 associated with $Pd_{1-x}Ag_x$ alloys.**

Overall, the EAM potentials provide an excellent representation of the structural properties of $Pd_{1-x}Ag_x$ alloys and a good agreement with the available elastic information. Further improvements in the elastic constants would require either additional experimental data for the larger silver concentrations, or a reformulation of the elemental potentials to provide a better agreement with the absolute values of the elastic constants. Both of these are beyond the scope of this work.

## 7. Characteristics of Palladium Silver Hydrides Compounds
### 7.1. Reported properties

Experimental works on $Pd_{1-x}Ag_xH_y$ compounds have predominately studied how additions of silver affect the hydrogen solubility. At room temperature and ambient pressure, a miscibility gap is observed with the absorption of hydrogen into pure palladium (x = 0) resulting in an α phase at y = 0.03 and a β phase with y = 0.60 [53]. Both phases are associated with a face-centered-cubic metal lattice with the hydrogen atoms occupying octahedral interstitial sites. In contrast, the solubility of hydrogen into silver is negligible [54]. Adding silver to palladium results in a decrease in both the width of the miscibility gap and hydrogen solubility, with the miscibility gap disappearing for silver concentrations greater than approximately x = 0.25 [7, 8].

Little to no experimental information has been obtained directly for the structural and elastic properties of the alloy hydrides. There is no clear evidence as to whether the hydrogen atoms prefer to occupy octahedral or tetrahedral interstitial sites. Speculation typically points to the octahedral site being favored as hydrogen occupies the octahedral site in pure palladium and other face-centered-cubic elements. However, neutron diffraction data on palladium-gold alloys infused with deuterium revealed the deuterium to occupy both types of interstitial sites, with the tetrahedral occupancy increasing with deuterium concentration [55]. This indicates the possibility that alloying may allow tetrahedral sites to be occupied. No measurements of the elastic properties have been found.

Insights into the structural properties of $Pd_{1-x}Ag_xH_y$ compounds have been provided by DFT calculations and experimental measurements on diffusivity. Previous DFT analyses predict that the hydrogen favors octahedral interstitial sites over tetrahedral sites for low concentrations of hydrogen and silver [51, 56]. Additionally, the hydrogen is observed to highly favor palladium atoms over silver atoms in that hydrogen located at palladium rich octahedral sites are energetically favorable to silver rich octahedral sites.

This preference of the palladium rich sites is supported by experimental diffusion data. The diffusivity of pure palladium and silver are comparable, but the diffusivity strongly decreases for the alloys [57, 58]. A simple diffusion model based on having favorable palladium rich sites and unfavorable silver rich sites was able to accurately reproduce the concentration dependence of the diffusivity. According to the model, as the silver content increases, diffusion decreases as favorable sites for the hydrogen become less prevalent. Once the alloys become predominately silver, the hydrogen becomes trapped at sites near palladium atoms resulting in further decreases in the diffusion rate.

### 7.2. Calculations of $Pd_{0.75}Ag_{0.25}H_y$

In addition to providing direct fitting information, the ideal $Pd_{0.75}Ag_{0.25}H_y$ systems studied with DFT provide a simplified structural representation. Eighteen unique $Pd_{0.75}Ag_{0.25}H_y$ systems are identified and investigated consisting of combinations of O1 and O2 sites, or arrangements of T sites to form compositions of y = 0.25, 0.5, 0.75 and 1. DFT results for y = 0.25 show that the O1 site is the most energetically favored and the O2 site is the least favored (table 4). This shows that the hydrogen has a preference for palladium rich sites over sites that neighbor silver atoms. For y = 0.5, 0.75 and 1, the most favored arrangements are associated with the hydrogen occupying tetrahedral sites. This change in preference may in part be due to the fact that octahedral arrangements at the higher concentrations require some of the unfavorable O2 sites to be filled, as only one out of every four octahedral sites is an O1 site. It should be pointed out that the mixed O1-O2 systems are only slightly less favorable than the lowest of the tetrahedral systems for y = 0.5 and 0.75, with differences in energy of only 0.007 and 0.014 eV/atom respectively.

Table 4. Cohesive energies of the various $Pd_{0.75}Ag_{0.25}H_y$ systems with the H atoms positioned either at octahedral or tetrahedral sites. Note the preference for one of the octahedral sites at y = 0.25 and preference for tetrahedral arrangements at higher H concentrations.

| H content, y | H occupancy | DFT Energy (eV/atom) |
| --- | --- | --- |
| 0.00 |  | -3.655 |
| 0.25 | O1 | -3.396 |
|  | O2 | -3.337 |
|  | T | -3.373 |
| 0.50 | O1,O2 | -3.180 |
|  | O2,O2 | -3.147 |
|  | T | -3.171 to -3.188 |
| 0.75 | O1,O2,O2 | -3.035 |
|  | O2,O2,O2 | -2.944 |
|  | T | -3.025 to -3.049 |
| 1.00 | O1,O2,O2,O2 | -2.863 |
|  | T | -2.919 to -2.943 |

Evaluations of the same $Pd_{0.75}Ag_{0.25}H_y$ systems are also performed with the EAM potentials. Prior to any structural relaxations, the trends associated with the energies are consistent with the DFT results. For y = 0.25, the O1 site is the lowest in energy while the O2 site is highest. Also, for larger values of y, the lowest energy system is associated with a T site combination as opposed to an O1-O2 site combination.

Performing an atomic energy minimization on the y = 0.25 systems shows that there are local minimums associated with the hydrogen occupying the O1 and O2 sites, but not the T site as the hydrogen atoms relax to O1 sites. Annealing the systems with NVT integration at 100 K for 100 picoseconds causes hydrogen initially at O2 sites to shift to O1 sites. The conversion from O2 to O1 sites occurs completely for potential h but not for potential m. With potential m, the annealed system has two hydrogen atoms that appear close to T sites. Figure 7 reveals that the two hydrogen atoms are in fact in-between ideal O1 and T sites and result from the two hydrogen atoms vying for the same O1 site. In lattice coordinates, the two hydrogen atoms are both trying to occupy the O1 site at (0.5, 0.5, 0.5), but end up repulsing each other resulting in their actual positions at (0.32, 0.63, 0.68) and (0.66, 0.35, 0.32) which are between the O1 site and the T sites at (0.25, 0.75, 0.75) and (0.75, 0.25, 0.25). As these atoms cannot reach the octahedral site, nor are truly at tetrahedral sites, they are referred to here as forming a near-tetrahedral pairing.

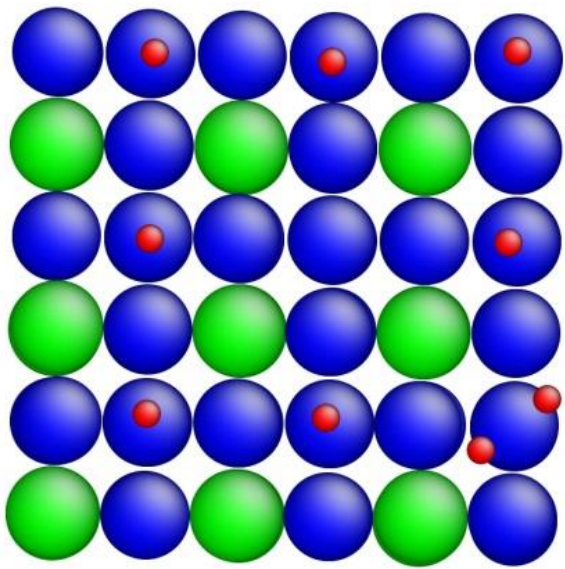

**Figure 7. <100> projection of an ideal $Pd_{0.75}Ag_{0.25}H_{0.25}$ system. Most of the hydrogen (small red) atoms occupy O1 octahedral sites directly above the Pd (blue) atoms, while two H atoms vying for the same O1 site form a near-tetrahedral pairing in the bottom right of the image.**

The preferred site occupancy is further explored for the $Pd_{0.75}Ag_{0.25}H_y$ by looking at simulations consisting of 108 metal atoms and 28 hydrogen atoms. As the favored O1 sites in these systems are fully saturated by 27 hydrogen atoms, the simulations investigate where the additional hydrogen atom prefers to be located. Three different variations are compared: one where the additional hydrogen atom is positioned at an O2 site, one where it is positioned at a T site, and one where it forms a near-tetrahedral pairing with one of the other hydrogen atoms. Annealing with potential m reveals the near-tetrahedral pairing to be preferred as the other two variations formed near-tetrahedral pairings. In contrast, potential h resulted in the additional hydrogen occupying an O2 site after annealing all three variations. The energy of the near-tetrahedral pairing system is found to be less than the O2 system by 0.11 eV for potential m and 0.01 eV for potential h. For comparison, DFT simulations of similar systems are also performed revealing the near-tetrahedral pairing having the lowest energy of the three, with a total cohesive energy of 0.06 eV less than the system with the O2 interstitial.

### 7.3. EAM predictions of $Pd_{1-x}Ag_xH_y$ compounds

Additional simulations using the EAM potentials are performed to evaluate the predicted properties of $Pd_{1-x}Ag_xH_y$ compounds ranging from x = 0 to 0.50 and y = 0 to 1.0. The hydride systems are built by adding hydrogen to the 4000 metal atom systems used to evaluate the $Pd_{1-x}Ag_x$ alloys. Hydrogen atoms are randomly inserted at octahedral interstitial sites in concentrations of every 0.1. The systems are annealed at 300 K and zero pressure using NPT integration for 50 picoseconds, before quenching to 0 K over 10 picoseconds and then relaxing the system by minimizing the atomic energy and forces.

A different, more complex occupancy behavior is observed for these larger systems than for the idealized systems studied in the previous section. The random silver positioning and varying compositions do not allow for the unique O1, O2, and T sites to exist. Rather, the tetrahedral and octahedral interstitial sites are associated with a wide variety of local environments. With both potentials, the hydrogen occupies both octahedral and tetrahedral interstitial positions for all systems containing silver. Plots of the fraction of hydrogen in tetrahedral sites are shown in figure 8, which were obtained by counting the number of hydrogen closer to tetrahedral sites than octahedral sites. Curves are not shown for $y \geq 0.8$ as the high hydrogen content causes local distortions that prevent the sites from being easy to distinguish. The fraction of tetrahedral hydrogen interstitials depends strongly on the silver concentration, ranging from practically no tetrahedral hydrogen in pure palladium to over 45% tetrahedral hydrogen for $x = 0.50$. Little dependence is seen on the site occupancy on the hydrogen concentration except for x near 0.50, where higher hydrogen concentrations, y, result in fewer tetrahedral hydrogen.

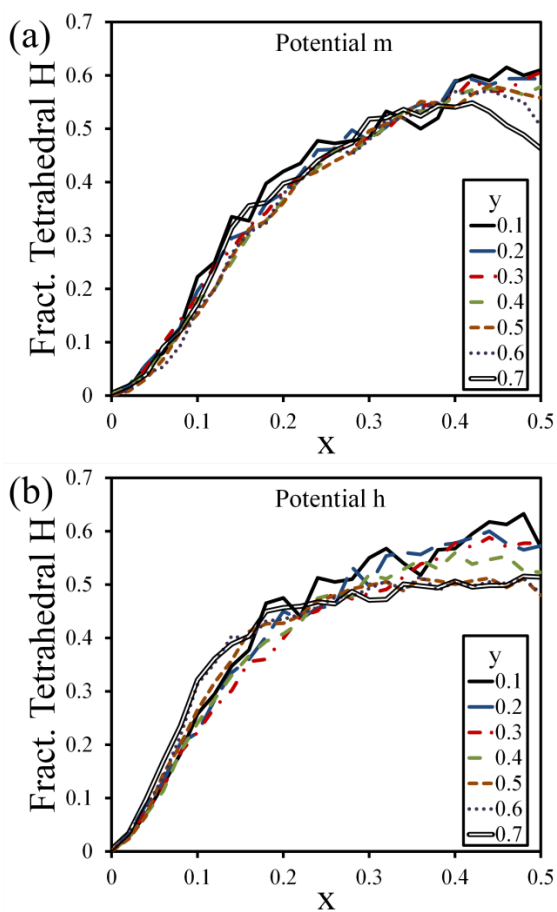

**Figure 8. Fraction of hydrogen atoms occupying tetrahedral sites after annealing $Pd_{1-x}Ag_xH_y$ systems with randomly positioned silver and hydrogen. Both potentials show a strong dependence on the silver concentration, x, and only a weak dependence on the hydrogen concentration, y.**

For $Pd_{0.98}Ag_{0.02}H_{0.2}$, 15 tetrahedral hydrogen atoms were found using potential m all at the ideal tetrahedral positions and not in near-tetrahedral pairings. Examinations of these 15 atoms reveal that none directly neighbor silver atoms, although all but one had silver atoms as next-nearest neighbors. As almost no tetrahedral hydrogen atoms are found for PdH$_y$, it is concluded that the long range interaction between silver and hydrogen is allowing for the tetrahedral sites to become stable favorable sites.

The lattice constant, cohesive energy and bulk modulus for Pd$_{1-x}$Ag$_x$H$_y$ compounds ranging from $0.00 \leq x \leq 0.50$ and $0.0 \leq y \leq 1.0$ are calculated with the EAM potentials, and the results are shown in figure 9. Smooth transitions of the properties are observed with changing either the silver or hydrogen compositions. Both EAM potentials have similar properties at low silver concentrations but show differences at higher concentrations. Most notably, potential h shows considerable elastic stiffening as the hydrogen content increases for compositions where $x \geq 0.30$ and $y \geq 0.4$.

Interestingly, the values measured here for PdH$_y$ with $y > 0.6$ deviate from the values reported by Zhou [12]. A close examination reveals the differences to be related to annealing the systems, with the high hydrogen annealed systems showing local lattice distortions due to the miscibility gap and hydrogen mobility. This annealing effect changes the bulk modulus the most by causing the values for $y > 0.6$ to remain nearly constant. A slight increase in the lattice parameter and decrease in potential energy are also observed to result from the annealing.

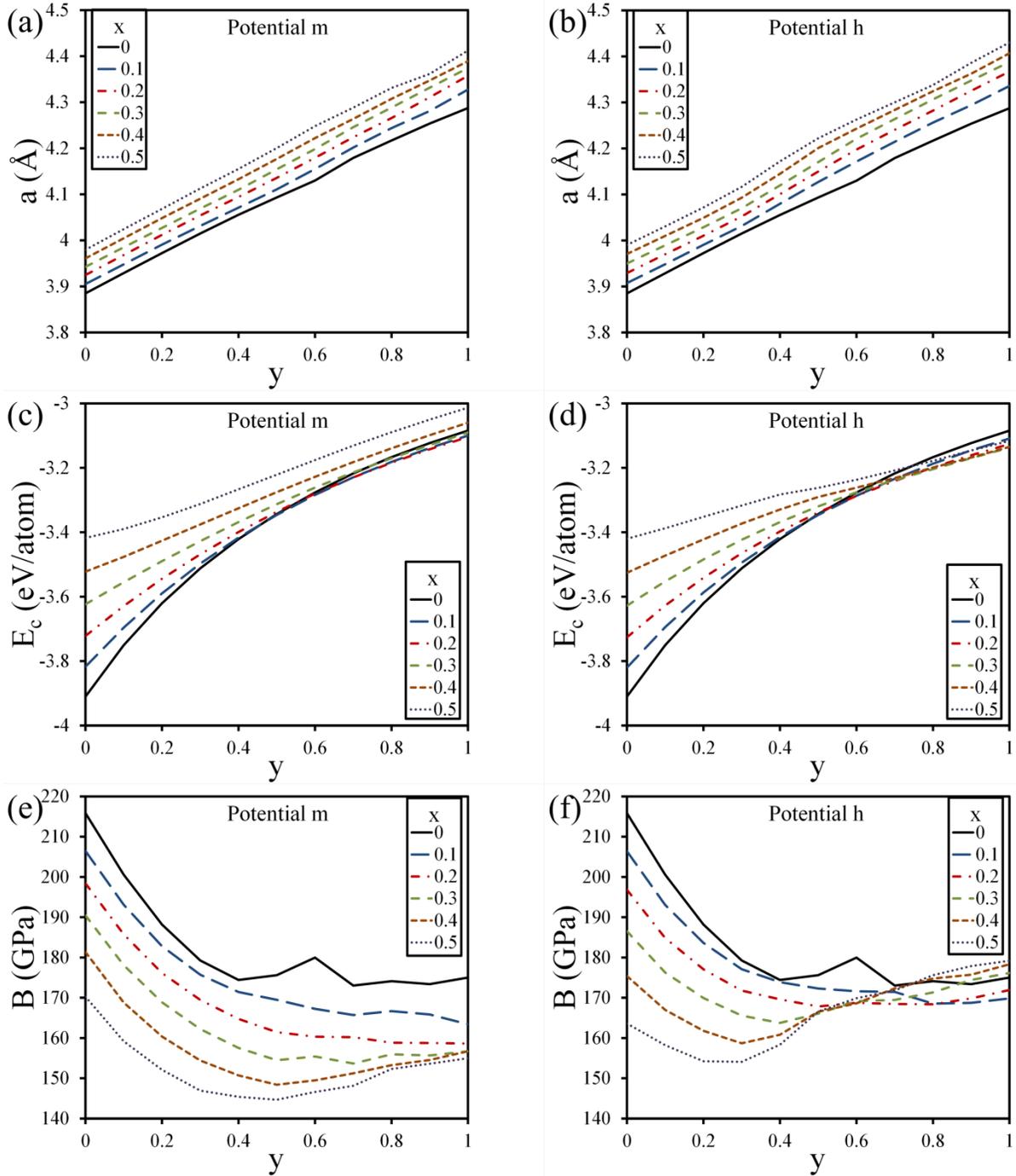

**Figure 9.** (a,b) Lattice constants, (c,d) cohesive energies and (e,f) bulk moduli for various compositions of $Pd_{1-x}Ag_xH_y$ as predicted by the EAM potentials.

Using the cohesive energy values, the change in energy associated with adding one additional hydrogen atom to the system is calculated and plotted in figure 10. This value is used here as a loose approximation of the hydrogen chemical potential. The earlier DFT works on $Pd_{1-x}Ag_xH_y$ noted that the adsorption energy associated with adding the first hydrogen into the metal alloys decreases with increasing silver content [51, 56]. This occurs as the silver expands the lattice making it easier for the hydrogen to fill the interstitial sites. Both EAM potentials show that increasing the silver content results in a larger energy drop associated with adding hydrogen into the pure metal alloys making the behavior of the potentials consistent with the DFT results.

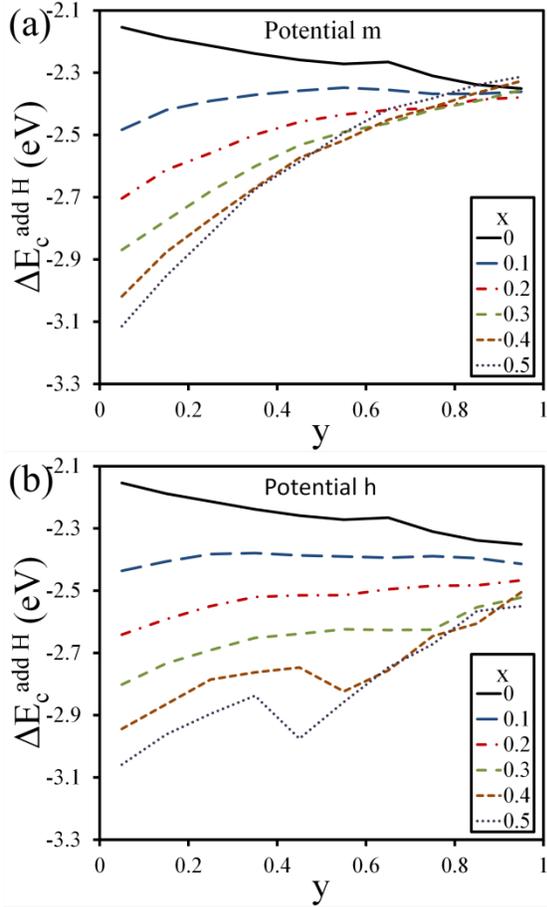

**Figure 10. The change in energy associated with adding a single hydrogen atom to various compositions of $Pd_{1-x}Ag_xH_y$.**

Additionally, these plots show a difference between the two potentials. Potential m shows smooth curves associated with the energy changes as a function of the hydrogen concentration, where the curves become steeper with increasing silver. In contrast, potential h has more level curves for x < 0.3, and sharp changes for the other compositions.

In order to explore how silver additions affects the miscibility gap of palladium hydride, the Gibbs free energy of mixing associated with adding hydrogen into a given metal composition is calculated. For this analysis, equations related to the ones derived by Zhou et al. [12] are used:

$$\Delta G^{mix} = \Delta H^{mix} - \Delta S^{mix} T \tag{15}$$

$$\Delta H^{mix} = E_{MH_y} - 2Y E_{MH} - (1-2Y) E_M \tag{16}$$

$$\Delta S^{mix} = -k_B \left[ Y \ln \frac{Y}{1-Y} + (1-2Y) \ln \frac{1-2Y}{1-Y} \right]. \tag{17}$$

Here, $Y$ is the mole fraction of hydrogen given by $Y = y/(1+y)$, $k_B$ is Boltzmann's constant, $E_M$ is the per atom cohesive energy of the hydrogen free alloy, $E_{MH}$ is the cohesive energy of the alloy with all octahedral sites filled with H, and $E_{MHy}$ is the cohesive energy associated with an intermediate concentration of H. For this calculation, the energy terms are taken without annealing the systems to ensure that the crystal systems remain true. It should be noted that these equations do not provide the total Gibbs free energy as they neglect the heat of mixing associated with Pd-Ag alloys and treat all sublattice sites equally during calculation of the entropy.

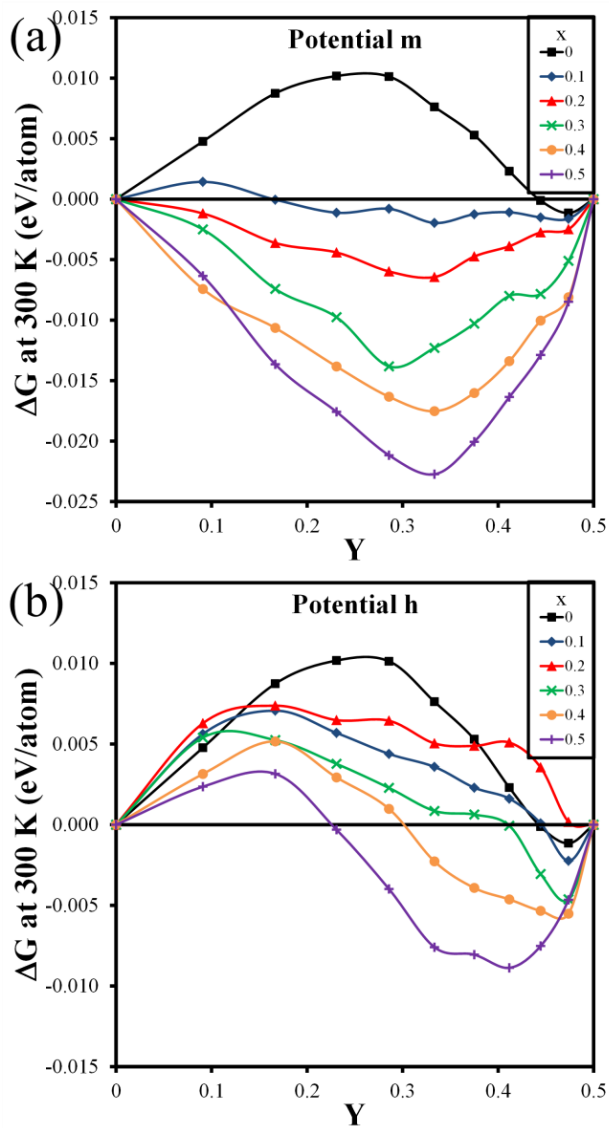

**Figure 11. The Gibbs free energy of mixing a mole fraction of hydrogen, Y = y/(1+y), into a given $Pd_{1-x}Ag_x$ alloy.**

The overall trends observed for the two potentials prove useful to the current analysis and display a clear difference between the two. A smooth curve is observed for $PdH_y$, with most compositions having positive Gibbs mixing energies, and a distinct minimum at Y = 0.47 associated with the known β phase. For potential h, adding silver does change the curves, but the overall trend remains similar, with the low hydrogen compounds having unfavorable positive Gibbs mixing energies, and some high hydrogen compounds having favorable negative energies. In contrast, adding silver with potential m results in all of the intermediate H compositions decreasing. By the time that x = 0.2 (for potential m), all of the calculated energies are negative indicating favorable mixing. These results suggest that potential m predicts the miscibility gap to disappear with increasing silver, while potential h predicts it to remain.

Considerable experimental evidence has investigated the pressure and phase stability of palladium silver hydrides, which can be directly compared to the Gibbs free energy results. The experimental results show that increasing the silver concentration causes the concentration of the a and b phases to shift closer to each other, eventually resulting in the loss of the miscibility gap for x = 0.25-0.30 [6-8]. The predictions for potential m therefore seem more in line with experiments than potential h, although a more thorough phase stability study is needed.

## 8. Conclusions

Two potentials for the palladium-silver-hydrogen system are presented here. For palladium-silver solid solution alloys, both potentials provide excellent representations of the heat of mixing and lattice parameters. The elastic constants for the metal alloys follow trends with composition that match with quantum calculations and experimental measurements. The potentials also show the general trends associated with the properties of the hydride, including the octahedral to tetrahedral site change suggested by quantum based results.

Both EAM potentials are presented here as they offer different representations of the hydrides, most notably in the Gibbs free energy of mixing and the nature of the mixed occupancy of the hydrogen interstitials. While the Gibbs free energy of mixing curves suggest that potential m is better at representing the loss of the miscibility gap with adding silver, the actual determination of which potential is better is left for future studies. Further evaluation of the potentials and the future development of improved potentials require additional measurements of the properties of both palladium-silver alloys and hydrides. Experimental measurements of the elastic properties would provide confidence in the values used for fitting. Neutron diffraction of palladium-silver-hydrides would provide a definitive answer on the hydrogen interstitial site occupancy. Further structural improvements can also be obtained through additional quantum calculations to investigate such things as long range interactions between silver and hydrogen, and the energy and structure of defects in the hydrides.

**Acknowledgements**

Sandia National Laboratories is a multi-program laboratory managed and operated by Sandia Corporation, a wholly owned subsidiary of Lockheed Martin Corporation, for the U.S. Department of Energy's National Nuclear Security Administration under contract DE-AC04-94AL85000.